\documentclass[a4paper,11pt]{article}
\usepackage{verbatim}
\usepackage{jinstpub} 
\usepackage{graphicx}

\usepackage{subcaption} 
\usepackage{hyperref}
\usepackage{booktabs}
\usepackage{makecell} 

\title{\boldmath Water Cherenkov detectors with fiber enhanced PMT for cosmic ray observation}

\author{H. Sun, Z. Huang, B. Wang, D. Liu, S. Ji,}
\author[1]{C. Feng\note{Corresponding author.}}
\affiliation{Key Laboratory of Particle Physics and Particle Irradiation (MOE)
, Institute of Frontier and Interdisciplinary Science, Shandong
University,\\
Qingdao, Shandong 266237, China}

\emailAdd{fengcf@sdu.edu.cn}

\abstract{Water Cherenkov detectors (WCDs) have been widely used in cosmic ray observations. This paper presents, for the first time, a cost-effective WCD design integrating a small photomultiplier tube (PMT) with wavelength-shifting fiber (WLS fiber) bundles. A WCD prototype was constructed in our laboratory, utilizing a fiber-PMT photosensor composed of a small PMT optically coupled with a WLS fiber bundle. This work details the structure of the fiber-PMT and the WCD. Measurements show the photosensor's time shift is 6.71 ns. Using cosmic ray muons, the WCD prototype demonstrated a light yield up to 30 photoelectrons with 2.3 ns time resolution under optimal conditions. Furthermore, tests under different trigger configurations indicate good performance uniformity. These findings validate the feasibility of the proposed design. A comparative analysis between a WCD using a standalone PMT and one equipped with the fiber-PMT shows that the fiber-PMT achieves nearly a 200\% improvement in light yield, albeit with a slight reduction in time performance.}


\begin{document}
\maketitle
\setcounter{footnote}{1}
\flushbottom

\section{Introduction}
\label{sec:intro}
Water Cherenkov detectors have been applied at several cosmic ray observatories, owing to their great performance and cost-effectiveness. Super-Kamiokande (Super-K) has the world’s largest Water Cherenkov detector (WCD) to detect neutrino interactions via Cherenkov light emitted by charged particles produced. The detector incorporates 11,146 Hamamatsu R3600 photomultiplier tubes (PMTs) with a diameter of 50 cm and 1,885 R1408 PMTs with a diameter of 20 cm~\cite{superK}. The High Altitude Water Cherenkov (HAWC) Observatory utilizes 300 WCDs in its main array, each equipped with one 10-inch PMT and three 8-inch PMTs to capture Cherenkov light generated by air showers from cosmic and gamma rays~\cite{hawc}. The Large High Altitude Air Shower Observatory (LHAASO) in China utilizes buried WCDs with 8-inch PMTs as muon detectors in its detector array. This WCD array uses PMTs of different sizes—8-inch, 20-inch, 3-inch, and 1.5-inch—to detect secondary particles in air showers~\cite{lhaaso}. Additionally, the forthcoming Southern Wide-field Gamma-ray Observatory (SWGO) also plans to use WCDs as its main detectors~\cite{swgo}. Large-sized PMTs are always preferred in WCDs to improve light collection and increase the experiment's sensitivity. They are the most expensive components of WCDs. However, if small-sized PMTs can achieve comparable light collection efficiency in WCDs, it could reduce the detector cost.

Wavelength-shifting fibers (WLS fibers) are a kind of special fibers doped with organic fluorescent material. It can absorb short-wavelength photons and re-emit light of longer wavelength. The re-emitted light then propagates through the fiber via total internal reflection. The Electromagnetic Detector (ED) in LHAASO uses a bundle of WLS fibers coupled with a small PMT to collect photons in a scintillator, achieving a time resolution of <~2~ns\cite{whitepaper}. A research team investigated the possibility of using WLS fibers in a WCD with a silicon photomultiplier and successfully separated cosmic muon signals from dark noise\cite{wcdSiPM}. Inspired by their work, a WCD design incorporating WLS fibers coupled with a small-sized PMT, akin to that of the ED in LHAASO, has been proposed. A detector prototype was built and its performance were tested.

\section{Design of Water Cherenkov Detector with Fiber-PMT}
\begin{figure}[htbp]
\centering
\begin{subfigure}[b]{0.31\textwidth}
    \centering
    \includegraphics[width=\textwidth]{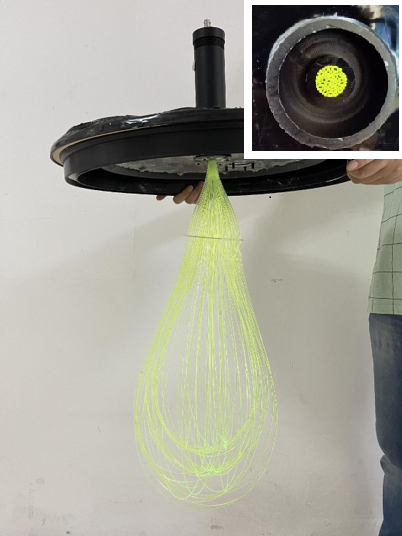}
    \caption{}
    \label{fig:fiberpmt}
\end{subfigure}
\quad
\begin{subfigure}[b]{0.31\textwidth}
    \centering
    \includegraphics[width=\textwidth]{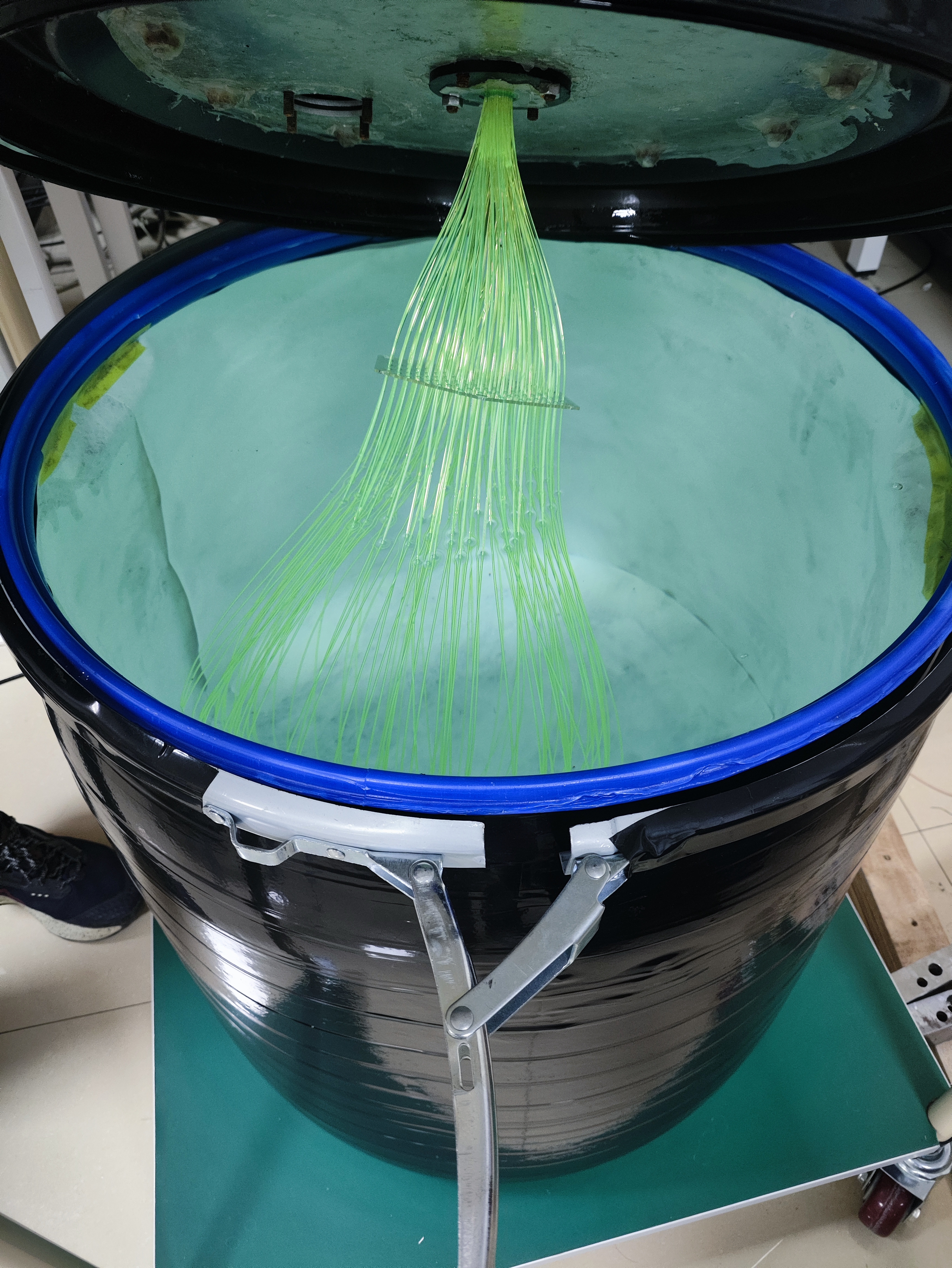}
    \caption{}
    \label{fig:tyvek}
\end{subfigure}
\quad
\begin{subfigure}[b]{0.31\textwidth}
    \centering
    \includegraphics[width=\textwidth]{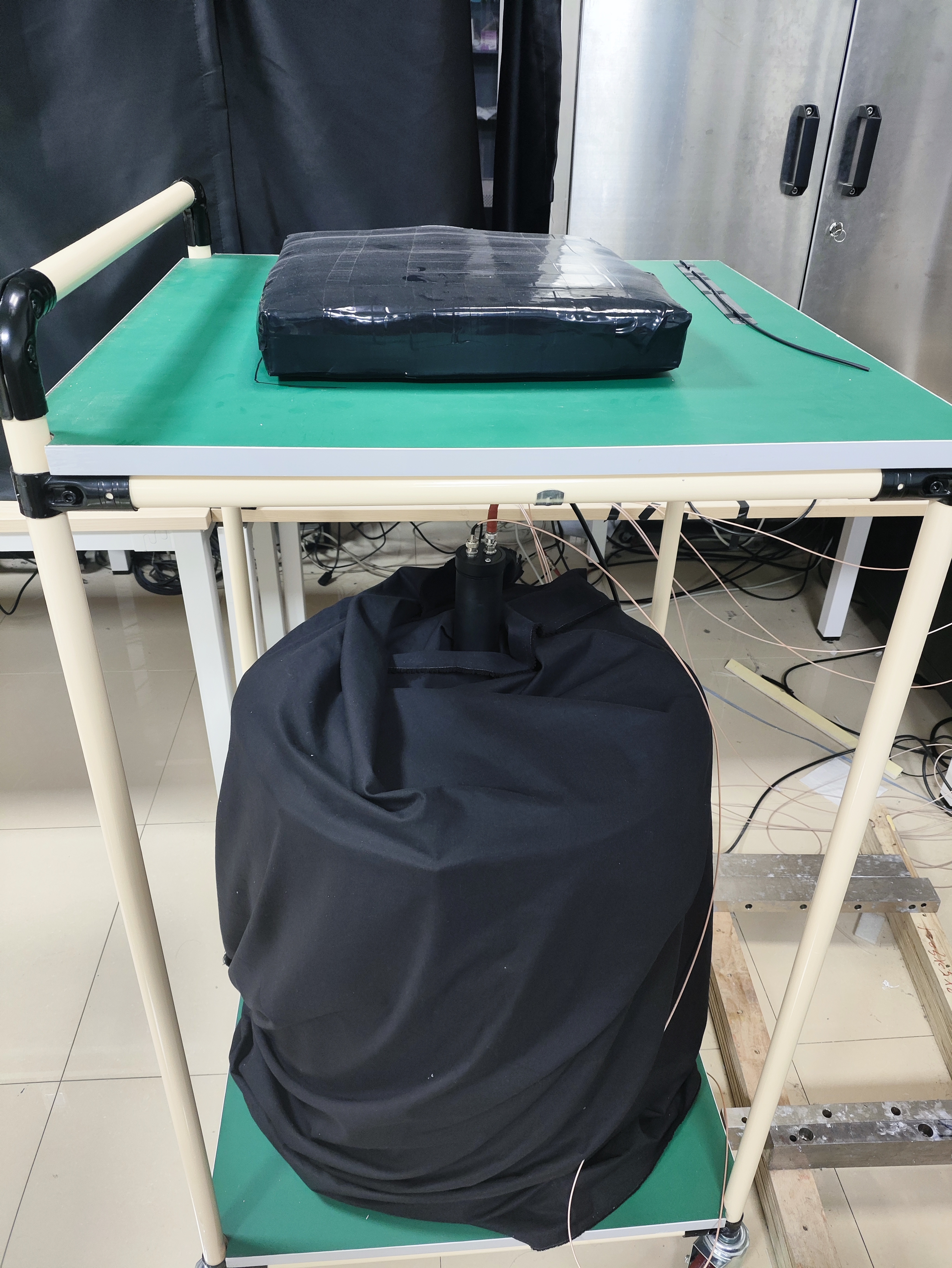}
    \caption{}
    \label{fig:detector}
\end{subfigure}
\quad
\caption{(a) Fiber-PMT assembly mounted on the tank lid. The top right corner shows the milled fiber ends. (b) The tank, filled with filtered water; the inner surface was lined with Tyvek 1085D, and the fibers are submerged. (c) Configuration of the experimental setup. Dual trigger detectors are positioned above and below the WCD.}
\end{figure}

To improve the light collection efficiency of small-sized PMTs in WCDs, we introduced a WLS-fiber-enhanced PMT, hereafter called fiber-PMT. Figure~\ref{fig:fiberpmt} displays an image of the fiber-PMT. The WLS fibers were bundled at the ends, while the middle part dispersed and was formed into a 'U' shape. The ends of fibers were milled and then connected to the PMT via the coupler.

The utilized PMT's model is XP3960 from HZC Photonics. It has a 1.5-inch concave, spherical alkali metal photocathode and a dynode chain consisting of 9 linear-focused stages. Additionally, it features a low dark noise count rate and high stability against temperature variations. Detailed specifications are listed in Table~\ref{tab:pmt}, and the quantum efficiency (Q.E.) curve is shown in Appendix~\ref{fig:QE-curve}. The utilized WLS fibers' model is Y-11 from Kuraray. Each fiber has a diameter of 1.2 mm and was trimmed to 1 meter in length before being bundled. They exhibit an emission peak at 476 nm and an absorption peak at 430 nm, with an attenuation length of approximately 3.5 meters according to the manufacturer's technical manual.

\begin{table}[h!]
\centering
\begin{tabular}{lc}
\hline
\textbf{Specifications} & \textbf{Values} \\
\hline
Photocathode size (inches) & 1.5 \\
Sensitive wavelength range (nm) & 300-700 \\
The gain ratio of anode and dynode & 112.6 \\
Maximum quantum efficiency (at 390 nm)& 25\% \\
Dark noise rate (Hz) & 2.05 \\
Temperature coefficient (\%/°C, from -30 °C to 30 °C) & -0.086 \\ 
\hline
\end{tabular}
\caption{Specifications of PMT XP3960\cite{yyh}}
\label{tab:pmt}
\end{table}

In the detector setup, a fiber-PMT was mounted on the lid of a cylindrical tank with diameter of 52 cm and height of 60 cm. The tank was filled with purified water, as shown in Figure~\ref{fig:tyvek}. The PMT was secured in a socket to prevent external light interference, and the fiber bundle was immersed in water to capture Cherenkov photons. This configuration enables external positioning of the PMT relative to the water tank, which improves maintenance accessibility. To further improve photon collection, Tyvek 1085D was used to line the inner surface of the tank as a diffuse reflector, with a reflectivity of 93.5\%, as measured by our lab's spectrum analyzer. Additionally, Additionally, the tank's exterior was wrapped in black tape and cloth to to block external light.

\newpage
\section{Performance Tests}
\subsection{Fiber-PMT time shift test}
\begin{figure}
\centering

\begin{subfigure}[b]{1\textwidth}
    \centering
    \includegraphics[width=\textwidth]{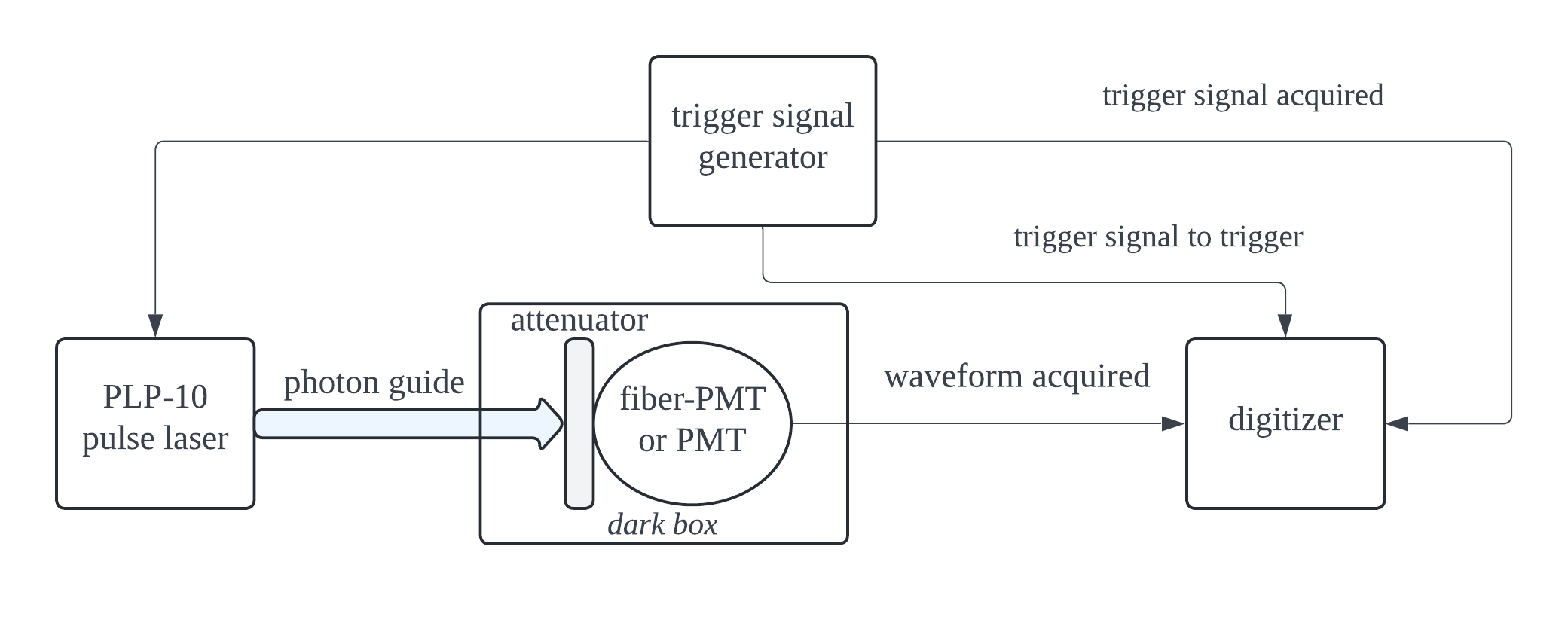}
    \caption{}
    \label{fig:timeshiftsetup}
\end{subfigure}

\begin{subfigure}[b]{0.325\textwidth}
    \centering
    \includegraphics[width=\textwidth]{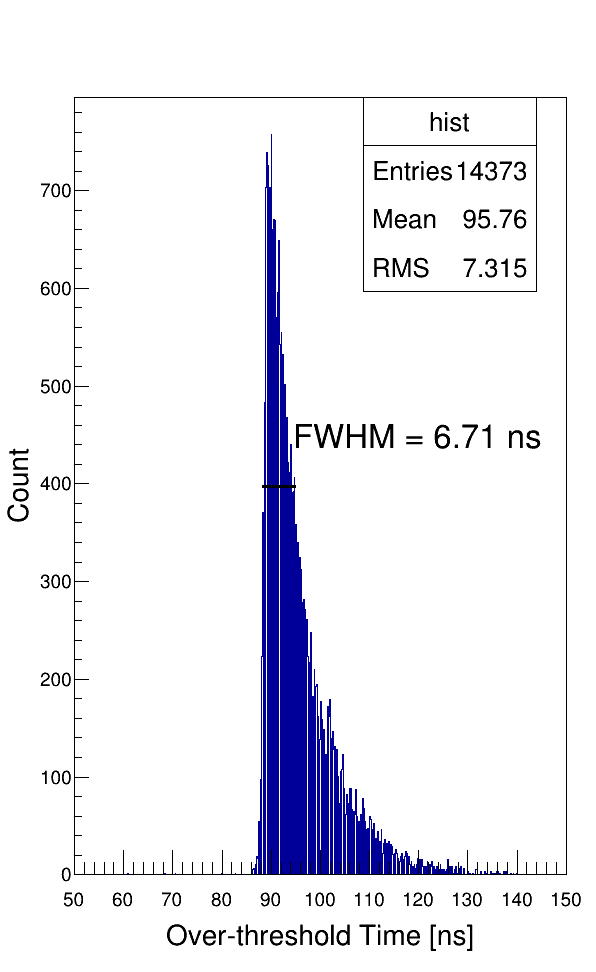}
    \caption{}
    \label{fig:singlephoton}
\end{subfigure}
\hspace{0em}
\begin{subfigure}[b]{0.325\textwidth}
    \centering
    \includegraphics[width=\textwidth]{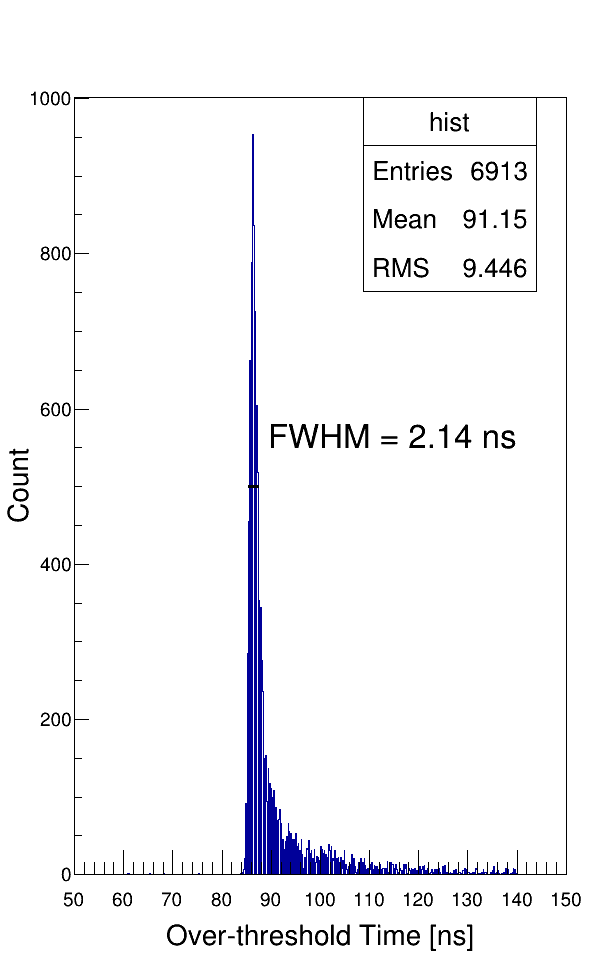}
    \caption{}
    \label{fig:PMT}
\end{subfigure}
\hspace{0em}
\begin{subfigure}[b]{0.325\textwidth}
    \centering
    \includegraphics[width=\textwidth]{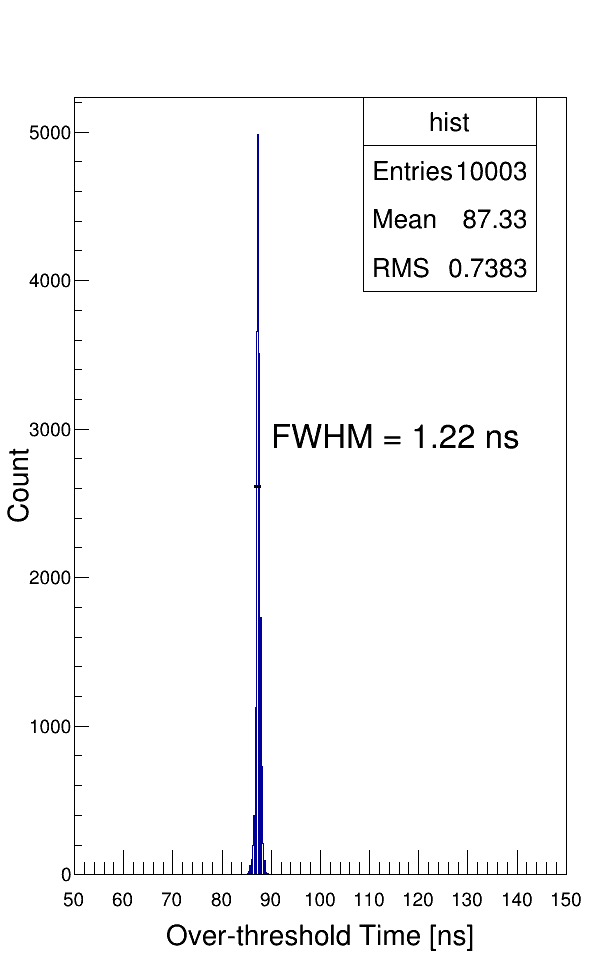}
    \caption{}
    \label{fig:multiphoton}
\end{subfigure}

\caption{(a) time shift test setup diagram. (b) Over-threshold time distribution for fiber-PMT on single photon mode. (c) Over-threshold time distribution for PMT on single photon mode. (d) Over-threshold time distribution for fiber-PMT on multi-photon mode.}
\label{fig:timeshiftresult}
\end{figure}

The time shift of the fiber-PMT was evaluated and compared with that of the standalone PMT. The test setup is depicted in Figure~\ref{fig:timeshiftsetup}. A signal generator produced three synchronized NIM signals: one triggered the pulse laser to emit photons, which illuminated either the fiber-PMT or the standalone PMT; another triggered the digitizer to capture the PMT's waveform signal; and the third was recorded by the digitizer as a reference signal to determine the time reference point. By synchronizing the NIM signals, the digitizer’s trigger timing was not used, fully avoiding its built-in timing jitter.
The laser system used for PMT and fiber illumination was a Hamamatsu PLP-10 picosecond pulse laser, emitting at 405 nm. The pulse width was precisely maintained at 70 picoseconds, with jitter kept below 10 picoseconds. Due to the high intensity of the laser, which hindered the PMT from operating in single-photon mode—an essential condition for accurate time shift measurements—a light attenuator was used to reduce the laser beam intensity, ensuring that the PMT could detect single photons.
The PMT signals were captured by the CAEN FADC V1743 digitizer, which features a sampling rate of 3.2GS/s, a 12-bit resolution, and an input dynamic range of 2.5 V$_{pp}$. The over-threshold time was defined as the moment when the waveform amplitude surpassed a predefined threshold of -5 mV. The full width at half maximum (FWHM) of the over-threshold time distribution was used to characterize the time shift performance. The measurement of the PMT in single-photon mode served as a reference, while the fiber-PMT was characterized in both single-photon and multiphoton modes by adjusting the laser intensity to obtain about 40 photons.

Figure~\ref{fig:singlephoton} illustrates the time shift measurement for the fiber-PMT in single-photon mode, with a FWHM value of 6.71 ns. Some bins appear empty due to the digitizer's limited sampling rate, making the data look more discrete. However, this does not affect the measurement accuracy, as the sampling rate remains sufficient to capture key timing information. The observed discreteness only impacts visualization, while the precision of time measurements is preserved. In comparison, Figure~\ref{fig:PMT} presents the measurement for the standalone PMT, which exhibits a smaller time shift of 2.14 ns. The larger time shift in the fiber-PMT is caused by the absorption and re-emission delay in the WLS fibers, as the final time shift consists of the contributions from fiber-PMT and the fibers. In multi-photon mode, the fiber-PMT exhibited a time shift of 1.22 ns (Figure~\ref{fig:multiphoton}), indicating its time response under typical detection conditions.

\subsection{Detector Performance Test}

\subsubsection{Experimental setup}

The performance of the WCD equipped with the fiber-PMT was evaluated using our cosmic ray test system, as illustrated in Figure~\ref{fig:detectorsetup}. Two trigger detector paddles were positioned above and below the WCD as shown in Figure~\ref{fig:detector}. As cosmic ray particles pass vertically through the system, both trigger detectors simultaneously produce signals. The signals are first processed by a discriminator set to a threshold of -10 mV, ensuring sufficient amplitude and effectively eliminating noise. Then the outputs of the two discriminators are sent to an AND logic gate. If they coincide, the logic gate generates a trigger signal to the V1743 digitizer, which then captures the signals from the detector. Several tests with different setups were conducted to evaluate the uniformity of the detector’s performance. In the vertical incidence setup, two large trigger paddles with size of 38 cm by 34 cm were placed vertically above and below the WCD. In the inclined incidence setup, the paddles were placed approximately diagonally. To map signal variations in smaller, specific areas of the detector, two small trigger detectors with size of 10 cm by 10 cm were used for central and peripheral incidences, placed vertically above and below the WCD, either at its center or at its periphery.
The self-trigger mode was also evaluated, in addition to tests using external triggers. In this setup, the detector’s signal was directly sent to the digitizer, which was operating in self-trigger mode. If a signal exceeded the predefined threshold of -7 mV, the digitizer automatically acquired the waveform.
\subsubsection{Data analysis}
The detector's performance, including light yield and time resolution, was measured based on signals captured by the digitizer. The number of photoelectrons (NPE) was used to represent the light yield and was determined by integrating the waveform for each event. The RMS value of the over-threshold time distribution served as a metric for time resolution, with the over-threshold time defined as the instant when the waveform amplitude exceeded the -5 mV threshold. The time at which the digitizer was triggered served as the reference, generally aligning with the timing of the two external triggers, although with a timing jitter of 2.9 ns. This timing jitter was subtracted from the RMS value of the histogram to determine the intrinsic time resolution of the detector.

\subsubsection{Test results}

\begin{figure}
\centering

\begin{subfigure}[b]{1\textwidth}
    \centering
    \includegraphics[width=\textwidth]{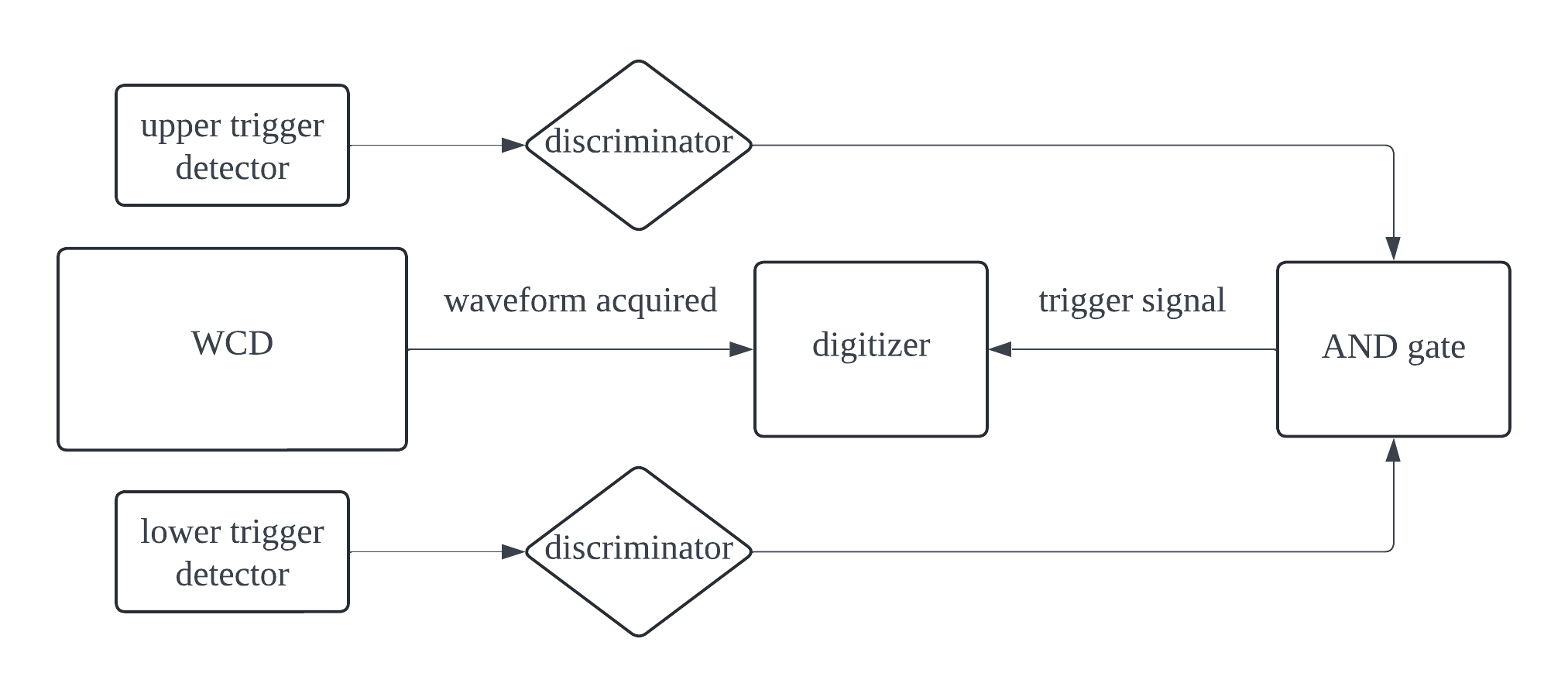}
    \caption{}
    \label{fig:detectorsetup}
\end{subfigure}
\begin{subfigure}[b]{0.49\textwidth}
    \centering
    \includegraphics[width=\textwidth]{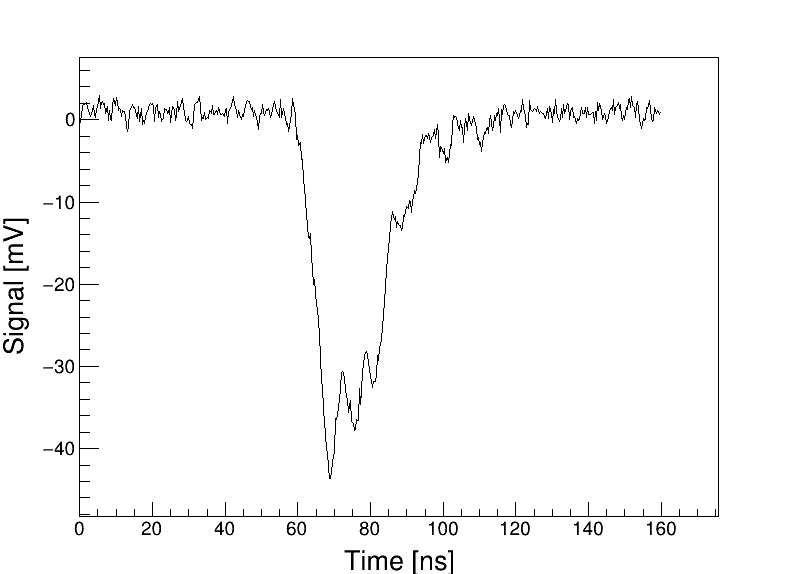}
    \caption{}
    \label{fig:waveform}
\end{subfigure}
\begin{subfigure}[b]{0.49\textwidth}
    \centering
    \includegraphics[width=\textwidth]{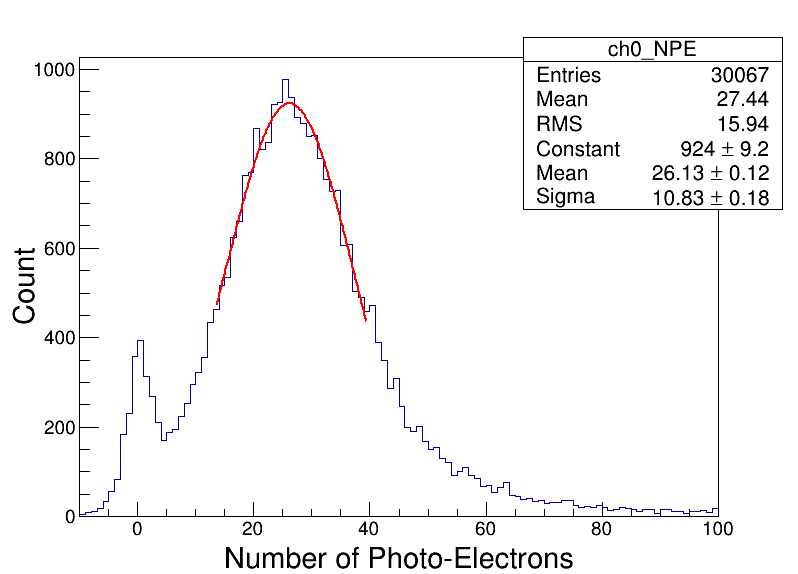}
    \caption{}
    \label{fig:verticalNPE}
\end{subfigure}
\hspace{1em}
\begin{subfigure}[b]{0.49\textwidth}
    \centering
    \includegraphics[width=\textwidth]{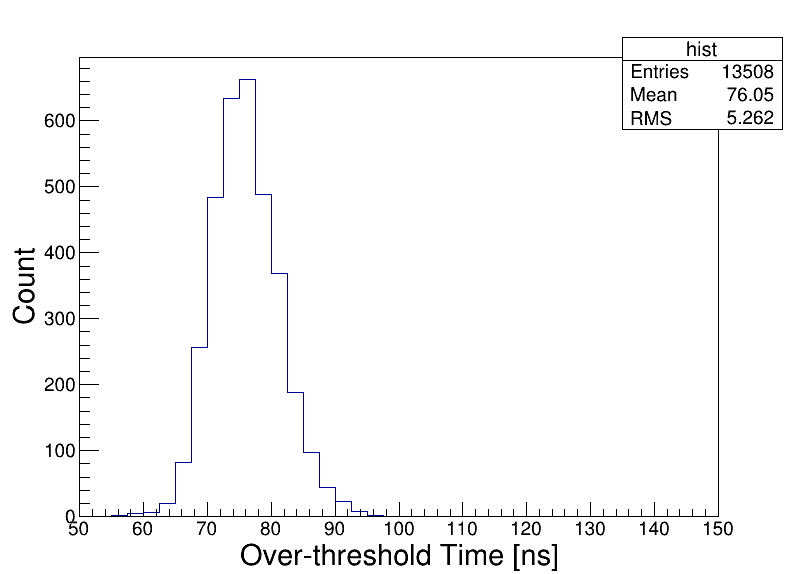}
    \caption{}
    \label{fig:verticaltime}
\end{subfigure}
\begin{subfigure}[b]{0.49\textwidth}
    \centering
    \includegraphics[width=\textwidth]{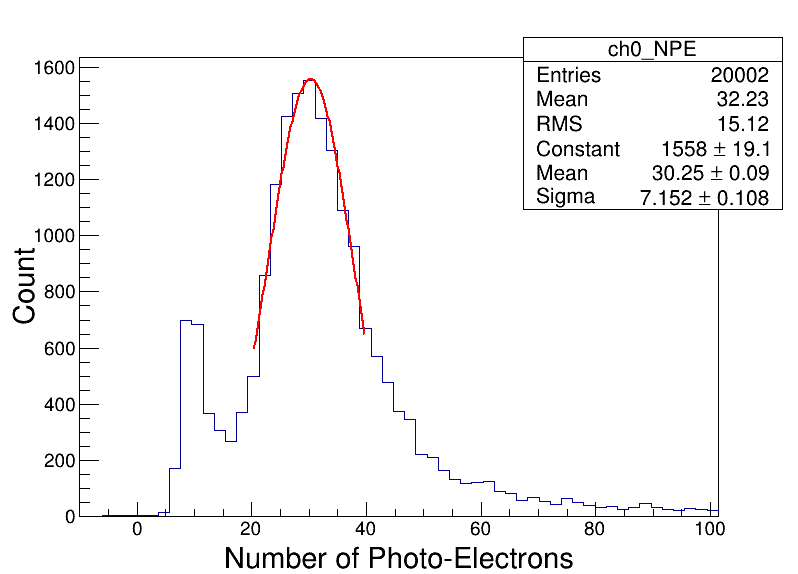}
    \caption{}
    \label{fig:self}
\end{subfigure}

\caption{(a) Detector performance test diagram. (b)~A waveform of WCD acquired by digitizer. (c)~NPE distribution under vertical incident trigger mode. (d)~Over-threshold time distribution under vertical incident trigger mode. (e)~NPE distribution under self trigger mode.}
\label{fig:vertical}
\end{figure}

Figure~\ref{fig:waveform} shows a sample waveform from the WCD as captured by the digitizer, characterized by a rapid rise of less than 10 ns and several minor peaks caused by photon reflections within the WCD. Typically, the complete signal lasts around 40 ns. Figure~\ref{fig:verticalNPE} displays the NPE distribution for vertical incidence. A minor peak near zero suggests the presence of non-signal events, or noise, likely caused by false triggering or detection failures. The prominent second peak, indicative of the actual signals from the WCD, was fitted with a Gaussian function. The valley between the non-signal peak and the signal peak follows the expected distribution pattern. The mean value from the Gaussian fit indicates a light yield of 26.13. Meanwhile, Figure~\ref{fig:verticaltime} displays the over-threshold time distribution, characterized by an RMS of 5.26 ns. After subtracting the time jitter of digitizer triggering, the intrinsic time resolution is 4.39 ns.

In the self-trigger condition, the time resolution is undefined due to the absence of a time reference, so we focus mainly on the light yield performance. Figure~\ref{fig:self} illustrates the NPE distribution in self-trigger mode. The results show a distribution pattern that resembles the one observed in the vertical incidence scenario with external triggers, characterized by two distinct peaks and a separation valley, corresponding to the noise peak and the complete signal peak. The mean NPE value is 30.25, slightly higher than that of the vertical incidence, likely due to the inclusion of incident particles from various directions, some of which produce larger signals due to increased track length. This result indicates that the WCD is capable of autonomously detecting muon signals without the need for an external trigger, which is particularly useful in scenarios where external triggers are unavailable.

The results from other trigger setups, together with the findings mentioned above, are summarized in Table~\ref{tab:summary1}. For a more detailed overview of all test results in various trigger setups, refer to the supplementary diagrams in Appendix~\ref{appendix}. Across various incidence scenarios, the WCD maintains a stable response, with an average light yield of approximately 30 photoelectrons, highlighting its detection reliability. 

In a more detailed analysis, the NPE increases by approximately 8.5\% in inclined incidences compared to vertical ones. This is likely because muons traveling a longer path, thereby generating more Cherenkov photons. Despite this increase, the time resolution remains consistent in both conditions, indicating the robustness of the detector to changes in the incident angle. In scenarios involving small trigger paddles, the central incidence produces a 23\% larger NPE value than the peripheral incidence, suggesting that the center of the WCD captures Cherenkov photons more effectively than the periphery. Additionally, the time resolution at the center is slightly better than that at the periphery. This may be due to the difference in photoelectron yield, since a larger number of photoelectrons results in a smaller time shift.

When comparing scenarios with large and small paddles, those with small trigger paddles exhibit better time performance. This is likely because large paddles cover a wider range of angles and tracks, increasing the complexity of photon propagation as WLS fibers are involved, which also broadens the NPE distribution (higher sigma value). Additionally, large trigger paddles have a greater inherent time shift due to their larger size, which results in photons being emitted from multiple positions on the paddle.

The self-trigger mode unexpectedly yielded a narrower NPE distribution (lower sigma value) than the external large-paddle triggers. This apparent anomaly arises from the self-trigger’s amplitude-based threshold: signals with low amplitude but wide pulse width—corresponding to high NPE values—are rejected, thereby eliminating part of the distribution and artificially reducing its sigma value. In contrast, external triggers could preserve the full NPE spectrum of the detector, resulting in a broader and more representative distribution.

\begin{table}
\centering
\caption{Summary of performance test results of detector with fiber-PMT.}
\begin{tabular}{cccccc}
\hline
Trigger Condition & \makecell{Avg. Track \\ Length [cm]} & Mean NPE & Sigma of fit & Time Resolution [ns] \\
\hline
Vertical(large paddles)  & 60 & 26.12 ± 0.12 & 10.83 ± 0.18 & 4.39 \\
Inclined(large paddles)  & 64 & 28.34 ± 0.17 & 12.18 ± 0.22 & 4.39 \\
Central(small paddles)  & 60 & 31.72 ± 0.37 & 7.15 ± 0.49 & 2.30 \\
Peripheral(small paddles)  & 60 & 25.84 ± 0.23 & 6.33 ± 0.33 & 2.61 \\
Self-trigger  & - & 30.25 ± 0.09 & 7.15 ± 0.11 & - \\  
\hline
\end{tabular}
\label{tab:summary1}

\end{table}

\subsubsection{WCD test without fibers}

\begin{figure}
\centering

\begin{subfigure}[b]{0.45\textwidth}
    \centering
    \includegraphics[width=\textwidth]{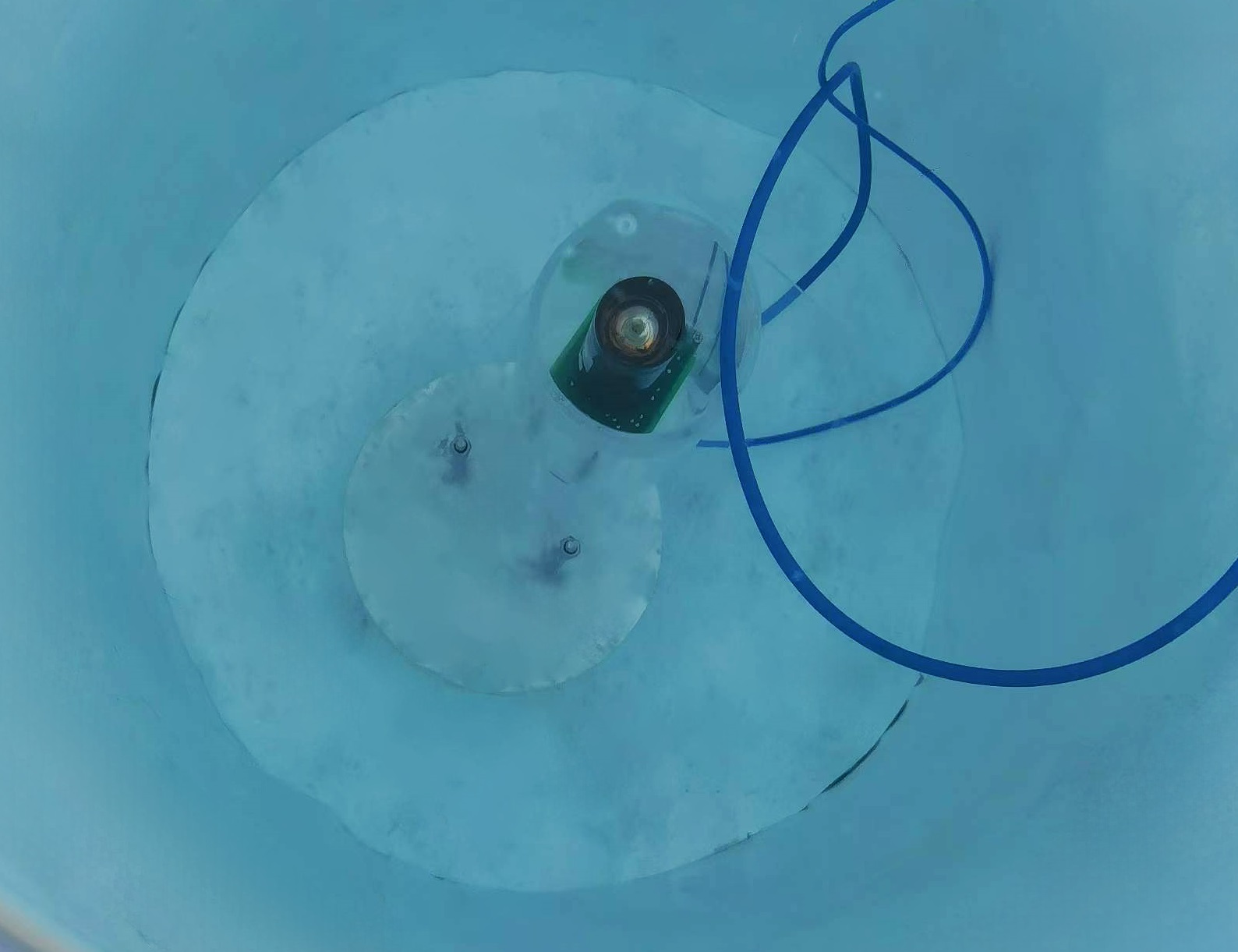}
    \caption{}
    \label{fig:PMTWCD}
\end{subfigure}
\begin{subfigure}[b]{0.49\textwidth}
    \centering
    \includegraphics[width=\textwidth]{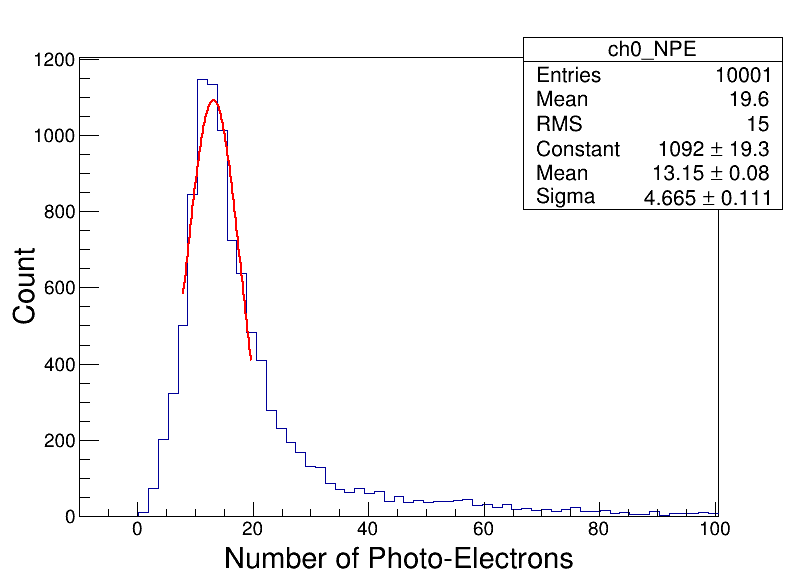}
    \caption{}
    \label{fig:PMTselfNPE}
\end{subfigure}
\begin{subfigure}[b]{0.49\textwidth}
    \centering
    \includegraphics[width=\textwidth]{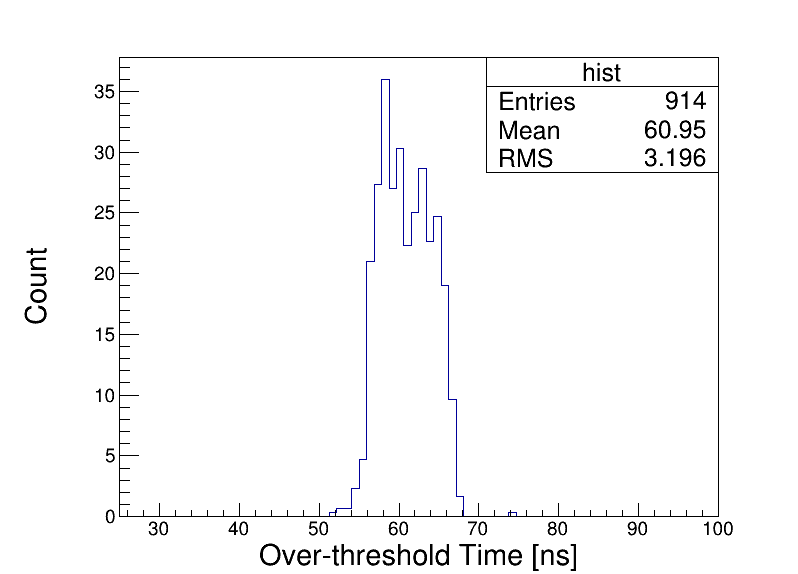}
    \caption{}
    \label{fig:PMTsmallcentraltime}
\end{subfigure}
\begin{subfigure}[b]{0.49\textwidth}
    \centering
    \includegraphics[width=\textwidth]{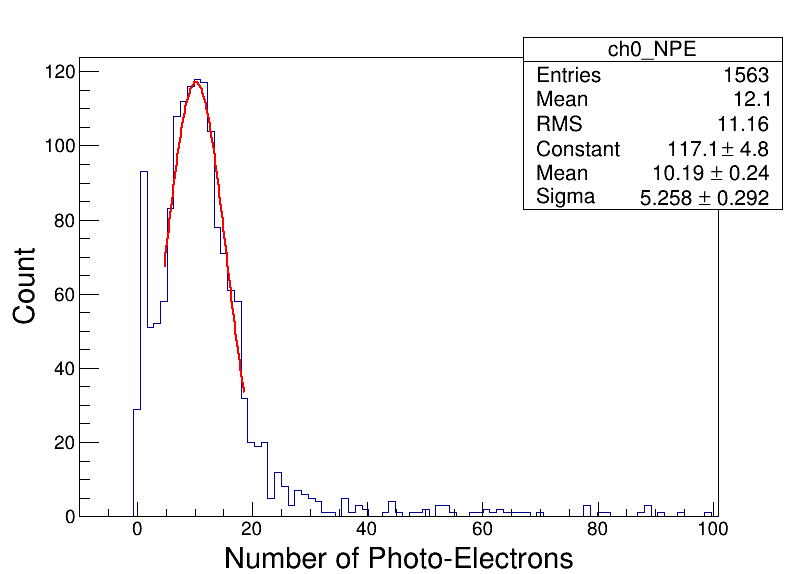}
    \caption{}
    \label{fig:PMTsmallcentralNPE}
\end{subfigure}
\begin{subfigure}[b]{0.48\textwidth}
    \centering
    \includegraphics[width=\textwidth]{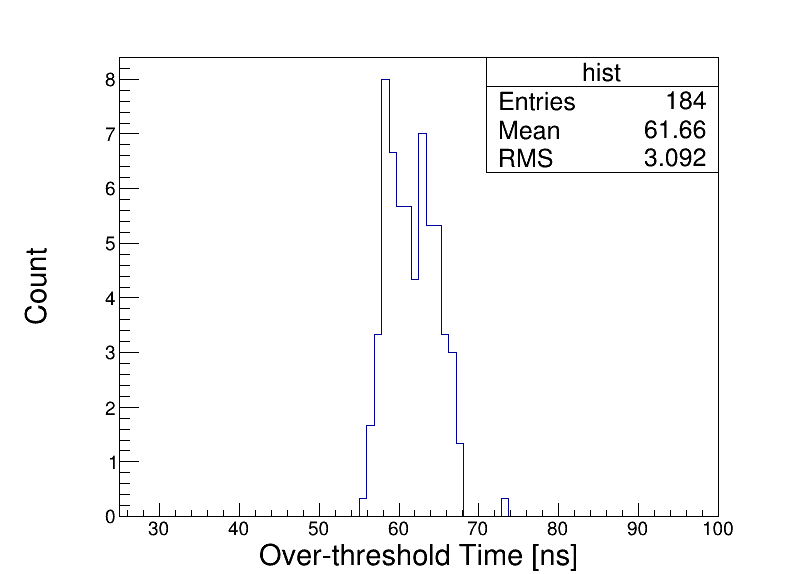}
    \caption{}
    \label{fig:PMTsmallperiphrealtime}
\end{subfigure}
\begin{subfigure}[b]{0.48\textwidth}
    \centering
    \includegraphics[width=\textwidth]{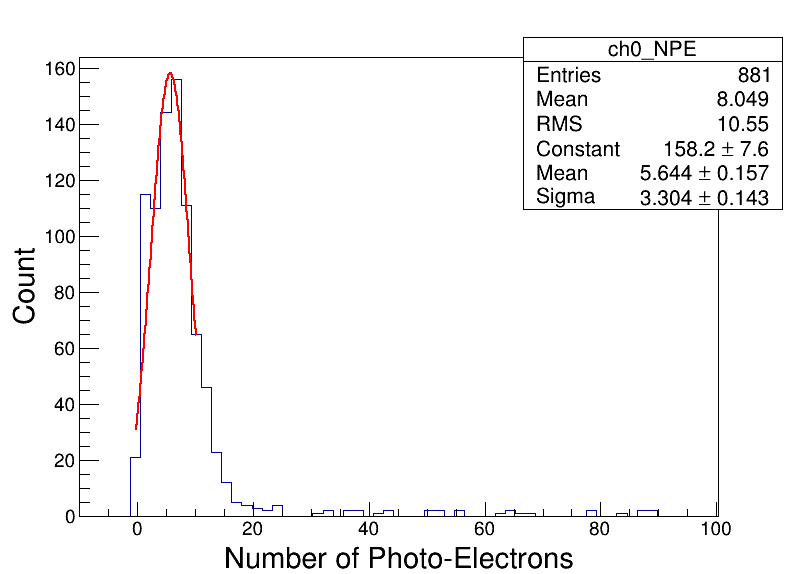}
    \caption{}
    \label{fig:PMTsmallperiphrealNPE}
\end{subfigure}
\caption{(a) PMT in the water-proof transparent case at the bottom of the tank. (b)~NPE distribution under self trigger mode. (c)~Over-threshold time distribution of central incidence. (d)~NPE distribution of central incidence. (e)~Over-threshold time distribution of peripheral incidence. (f)~NPE distribution of peripheral incidence.}
\label{fig:vertical}
\end{figure}

The performance of the detector without the fibers was also evaluated for comparison, providing a baseline to assess the influence on the detector brought by the fibers. Given the requirement for the Cherenkov light to meet a specific acceptance angle for detection by the photon sensor, we positioned the PMT at the center bottom of the tank to ensure optimal light collection. To achieve this, we designed a waterproof transparent casing (with transparency of about 92\%) to house the PMT, ensuring it remained isolated from direct contact with water, as illustrated in Figure~\ref{fig:PMTWCD}. To maintain an equivalent effective water height for a fair comparison with the fiber-PMT setup, the PMT was placed at the bottom of a taller tank. Three setups were tested: one with central incidence, one with peripheral incidence, both using small trigger paddles, and a third with a self-trigger setup.

Figures~\ref{fig:PMTsmallcentralNPE} and \ref{fig:PMTsmallcentraltime} show the NPE distribution and over-threshold time distribution for central incidence. The measured light yield was approximately 10, with a time resolution of 1.34 ns after correcting for the digitizer’s timing jitter. In contrast, peripheral incidence exhibited a 45\% reduction in light yield, see~\ref{fig:PMTsmallperiphrealNPE}. While the time performance at the periphery is slightly better as~\ref{fig:PMTsmallperiphrealtime} shows. This is because the predefined threshold for timing remains the same in all scenarios, and the peripheral scenario has smaller signal amplitudes. As a result, the threshold filters out many smaller signals, allowing only the larger ones to pass through, which ultimately improves time performance. The result might be affected by an incomplete data set, leading to potential bias.

The fiber-PMT WCD setup demonstrated a significant improvement, achieving nearly 200\% higher light yield than the PMT setup. This enhancement is due to the WLS fibers redirecting Cherenkov photons toward the PMT through absorption and re-emission, which expands the active detection area. Notably, the fiber’s photon collection efficiency remained high even for peripheral incidences. However, this advantage comes with a trade-off in time resolution. The degraded time performance of the fiber-PMT setup can be attributed to an increased time shift caused by absorption and re-emission processes in the WLS fibers. Conversely, the higher photoelectron yield partially compensates for this time shift by improving signal statistics. Ultimately, despite these opposing effects, the detector’s overall time resolution remains comparable to that of the PMT alone.

Figure~\ref{fig:PMTselfNPE} shows the NPE distribution for PMT-only setup in self-trigger mode. The light yield is around 13, 56\% lower than that of the fiber-PMT. The single muon peak lacks a well-defined separation due to the absence of a distinct valley, suggesting that the WCD with PMT alone cannot observe the single muon peak, whereas the WCD with fiber-PMT can. The results are summarized in Table~\ref{tab:summary2}.

The use of WLS fibers in the detector design leads to a substantial increase in light yield, significantly improving photon collection efficiency over small PMTs alone. Although the time resolution decreases modestly, it can be further improved with optimizations of the fiber-PMT design in the future. Overall, the fiber-PMT setup provides a highly viable and cost-effective solution for Water Cherenkov detectors.

\begin{table}
\centering
\caption{Summary of performance test results of detector with PMT.}
\begin{tabular}{ccccc}
\hline
Trigger Condition & \makecell{Avg. Track \\ Length [cm]} & Mean NPE & Sigma of fit & Time Resolution [ns] \\
\hline
Central(small paddle) & 60 & 10.19 ± 0.24 & 5.26 ± 0.29 & 1.34 \\
Peripheral(small paddle) & 60 & 5.64 ± 0.16 & 3.30 ± 0.14 & 1.07 \\
Self-trigger & - & 13.15 ± 0.08 & 4.67 ± 0.11 & - \\  
\hline
\end{tabular}
\label{tab:summary2}

\end{table}

\section{Summary}
This study presents the first experimental demonstration of integrating fiber-PMT technology into WCD. Specifically, The time shift of the photosensor, and detector's performance like light yield, time resolution and uniformity were systematically evaluated. The fiber-PMT configuration achieved a light yield of approximately 30 photoelectrons per event, a 200\% improvement compared to PMT alone, with a slight trade-off in the time resolution. The detector demonstrated good performance uniformity over various trigger configurations, including self-trigger operation without external triggers. The integration of WLS fibers significantly enhanced photon collection efficiency, addressing a critical limitation of small PMT in water Cherenkov detectors. These results demonstrate that fiber-PMTs provide a practical and budget-friendly option for future water Cherenkov detectors. They can be easily employed for use in observatories while still maintain reliable detection performance.

It is worth mentioning that The WLS fibers used from our laboratory’s stock did not perfectly match the PMT photocathode’s sensitivity, slightly decreasing the number of photoelectrons. Using PMT and WLS fibers with better-matched emission wavelengths could further improve the detector's performance. Future work could focus on additional optimization of the fiber-PMT’s design, including adjustments to its shape and experimentation with the number, diameter, length, and type of fibers.

\acknowledgments
This work was supported by the National Key R\&D program of China (Grants No. 2024YFA1611404) and the National Natural Science Foundation of China (Grants No. 12175121).

\appendix
\section{Appendix: supplementary diagrams}  
\label{appendix}  

\begin{figure}[ht!]
    \centering
    \includegraphics[width=0.6\linewidth]{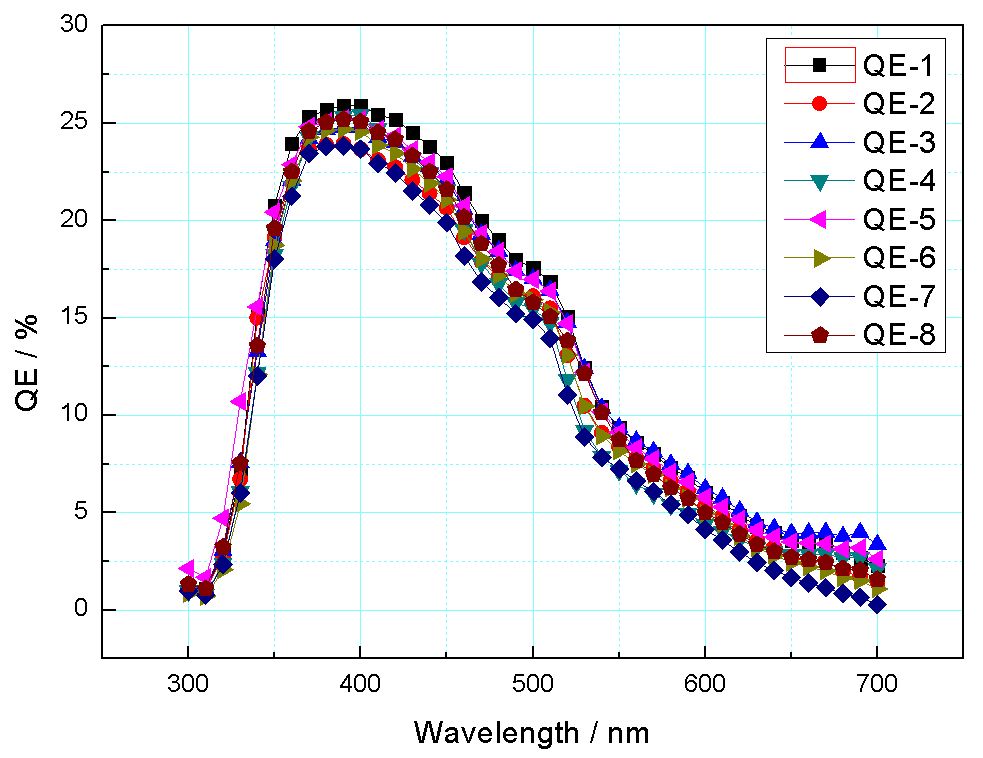}
    \caption{QE curve vs. wavelength of PMT XP3960 samples, tested in lab.}
    \label{fig:QE-curve}
\end{figure}

\newpage

\begin{figure}[ht!]
\centering

\begin{subfigure}[b]{0.48\textwidth}
    \centering
    \includegraphics[width=\textwidth]{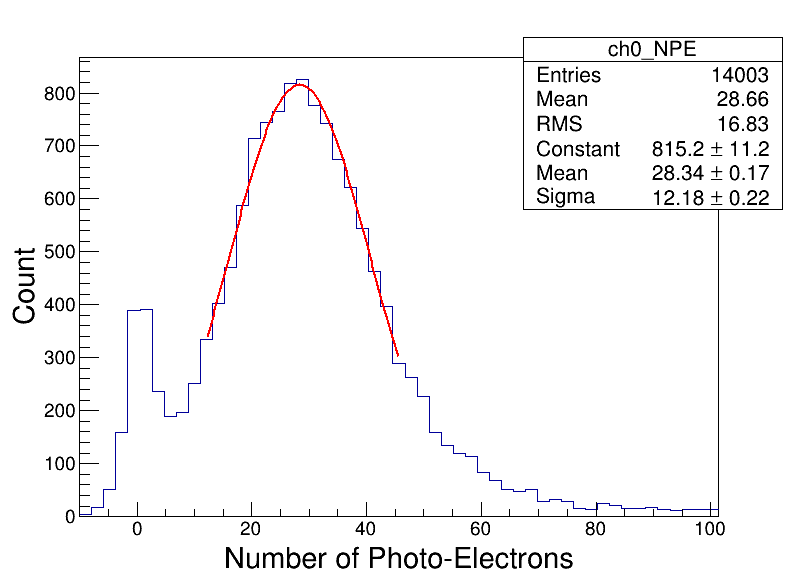}
    \caption{}
\end{subfigure}
\begin{subfigure}[b]{0.48\textwidth}
    \centering
    \includegraphics[width=\textwidth]{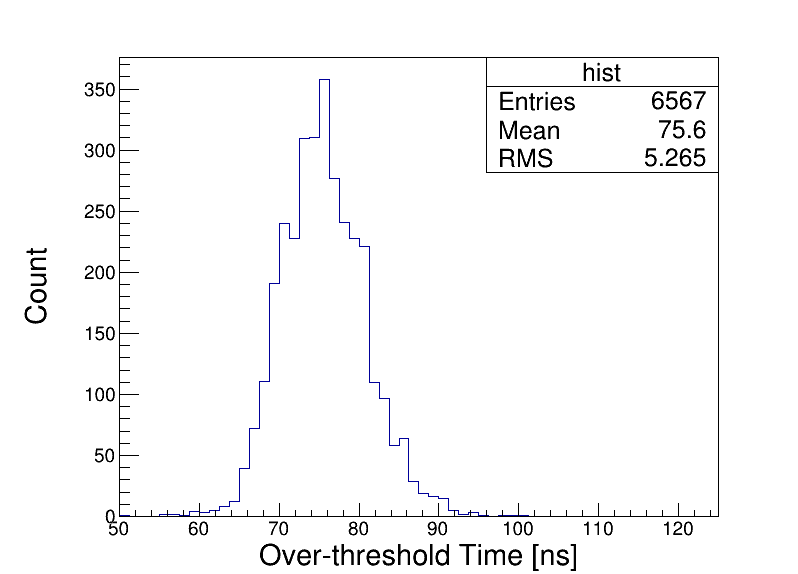}
    \caption{}
\end{subfigure}

\begin{subfigure}[b]{0.48\textwidth}
    \centering
    \includegraphics[width=\textwidth]{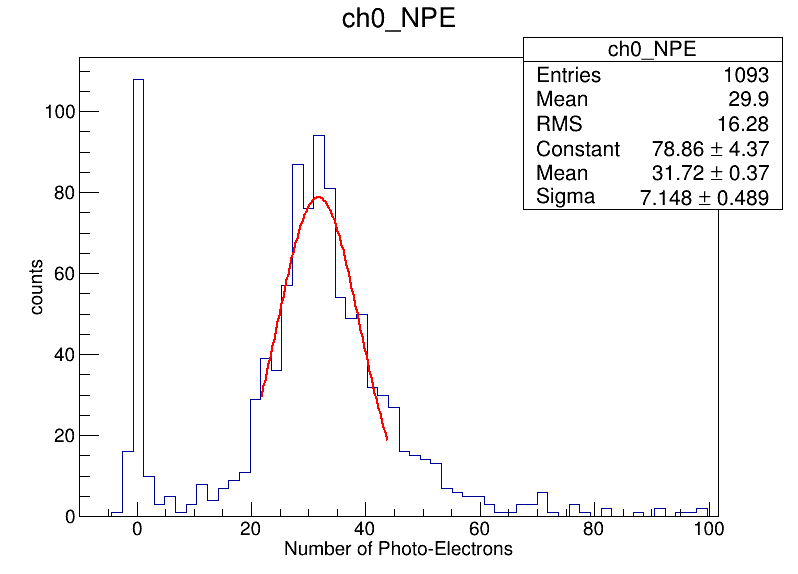}
    \caption{}
\end{subfigure}
\begin{subfigure}[b]{0.48\textwidth}
    \centering
    \includegraphics[width=\textwidth]{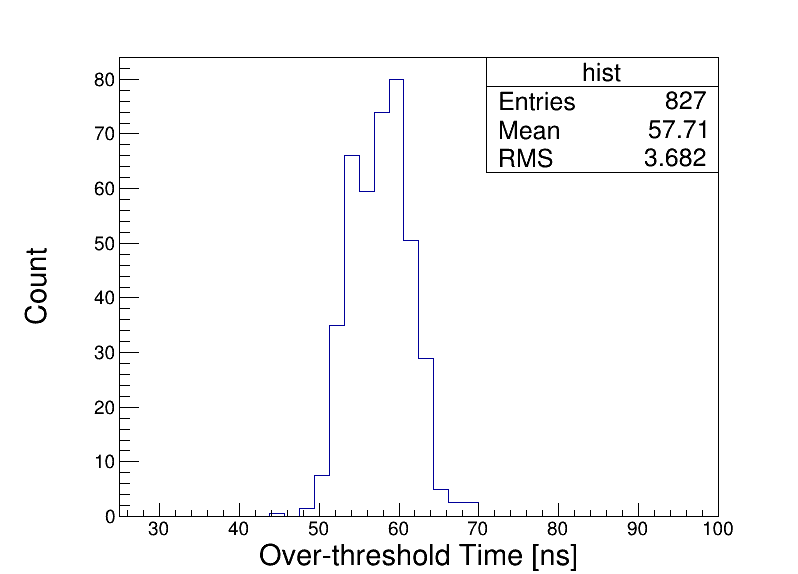}
    \caption{}
\end{subfigure}

\begin{subfigure}[b]{0.48\textwidth}
    \centering
    \includegraphics[width=\textwidth]{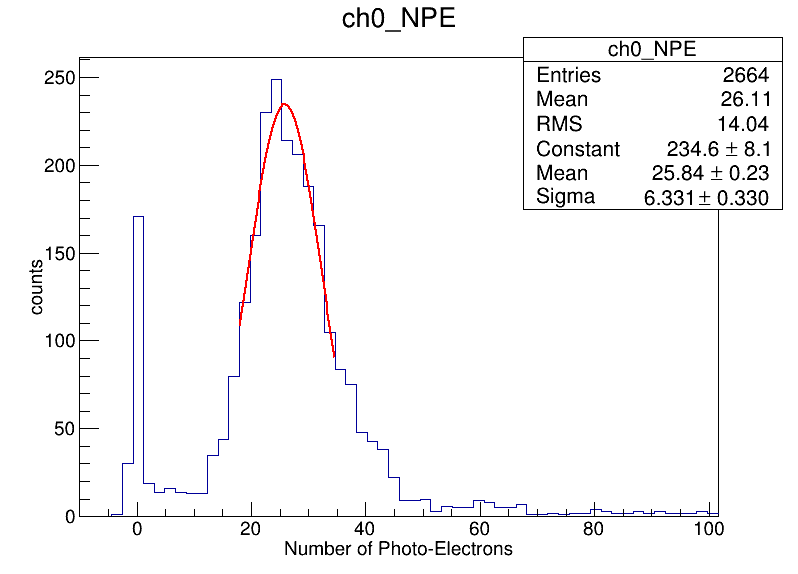}
    \caption{}
\end{subfigure}
\begin{subfigure}[b]{0.48\textwidth}
    \centering
    \includegraphics[width=\textwidth]{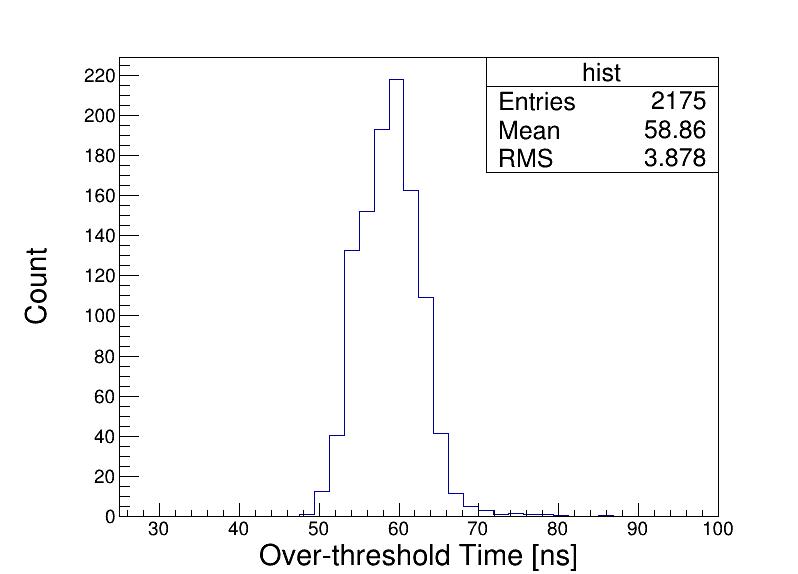}
    \caption{}
\end{subfigure}

\caption{Supplementary test results of WCD with fiber-PMT. (a) NPE distribution under inclined incident trigger mode. (b) Over-threshold time distribution under inclined incident trigger mode. (c) NPE distribution under central incident trigger mode. (d) Over-threshold time distribution under inclined central trigger mode. (e) NPE distribution under peripheral incident trigger mode. (f) Over-threshold time distribution under peripheral incident trigger mode. }
\label{fig:appendix}
\end{figure}

\end{document}